\documentclass{aastex6}

\fullcollaborationName{The Friends of AASTeX Collaboration}

\begin{document}

\title{Observations of Running penumbral waves emerging in a sunspot}

\author{T.G. Priya \altaffilmark{1,2},  Wenda Cao \altaffilmark{3},  Jiangtao Su\altaffilmark{1,2}, 
	 Jie Chen\altaffilmark{1}, Xinjie Mao\altaffilmark{1,4}, Yuanyong Deng\altaffilmark{1,2} and Robert Erd\'{e}lyi \altaffilmark{5,6}}
\affil{1. Key Laboratory of Solar Activity, National Astronomical
Observatories, Chinese Academy of Sciences, Beijing 100012, China\\
2. University of Chinese Academy of Sciences 19 A Yuquan Rd, Shijingshan District, Beijing 100049, China\\
3.Big Bear Solar Observatory, 40386 North Shore Lane, Big Bear City, CA 92314, USA\\
4.Beijing Normal University No. 19, XinJieKouWai St., Beijing 100875, China\\
5. Solar Physics and Space Plasma Research Centre, School of Mathematics and Statistics,\\ 
University of Sheffield, Hicks Building, Hounsfield Road,P.O.Box 32, Sheffield S3 7RH, UK\\
6. Department of Astronomy, E\"otv\"os Lor\'and University,P\'azm'any P. s\'et\'any 1/A, Budapest,H-1518, Hungary\\ }

\begin{abstract}

We present results from the investigation of  5-min umbral oscillations in a single-polarity sunspot of active region NOAA 12132. The spectra of TiO, H$\alpha$, and 304 \AA{} are used for corresponding atmospheric heights from the photosphere to lower corona.  
Power spectrum analysis at the formation height of H$\alpha$ - 0.6 \AA{} to H$\alpha$ center resulted in the detection of 5-min  oscillation signals in intensity interpreted as running waves outside the umbral center, mostly  with vertical magnetic field inclination $>15\degr$.
A phase-speed filter is used to extract the running wave signals with speed  $v_{ph}> 4$ km s$^{-1}$, from the time series of  H$\alpha$ - 0.4 \AA{} images, and found twenty-four 3-min umbral oscillatory events in a duration of one hour. 
Interestingly, the initial emergence of the 3-min umbral oscillatory events are noticed closer to or at umbral boundaries. These 3-min umbral oscillatory events are observed for the first time as propagating from a fraction of preceding Running Penumbral Waves (RPWs).
These fractional wavefronts rapidly separates from  RPWs and move towards umbral center, wherein they expand radially outwards suggesting the  beginning of a new umbral oscillatory event. 
We found that  most of these umbral oscillatory events develop further into RPWs. We speculate that the waveguides of running waves are twisted in spiral structures and hence the wavefronts are first seen at high latitudes of umbral boundaries and later at lower latitudes of the umbral center.

\end{abstract}

\keywords{Sun: sunspots-oscillation --- Sun: magnetic fields ---Sun: chromosphere}

\section{Introduction} \label{sec:intro}

 The first observational evidence of running penumbral waves
(RPWs) came from \citet{1972SoPh...27...71G} and \citet{1972ApJ...178L..85Z},
who detected concentric intensity waves propagating outward
through the penumbra of a sunspot in H$\alpha$ and having azimuthal
extents of $90\degr-180\degr$ and sometimes $360\degr$. These waves are considered to
be magnetoacoustic modes, were observed to propagate with a phase
velocity of 10$-$20 km s$^{-1}$ and exhibited intensity fluctuations
in the range of 10\% $-$ 20\%. \citet{1997ApJ...478..814B} and \citet{2004A&A...424..671K} have revealed how the frequencies
and phase speeds of RPWs vary from 3 mHz, 40 km s$^{-1}$  to 1 mHz, 10 km s$^{-1}$  from the inner penumbral boundary to outer penumbral edge, and becomes gradually invisible while approaching the outer boundary of the penumbra.
Additionally, Kobanov (2000) has observed the propagation of RPWs in the chromosphere up to $\sim15$$\arcsec$ ($\sim10,000$ km) from
the outer edge of the penumbral boundary, suggesting the Quiet Sun {\it p}-mode oscillations dominate at greater distances, hence
overpowering the signatures of any remaining RPWs.

The origin of RPWs has been under debate over years since their discovery, with current research attempting to address whether they are trans-sunspot waves of purely chromospheric origin (e.g., Tziotziou et al. 2006, 2007 and references therein) or the chromospheric signature of upwardly
propagating {\it p}-mode waves (Christopoulou et al. 2000, 2001;
Georgakilas et al. 2000; Centeno et al. 2006). \citet{2016ApJ...830L..17Z} have acknowledged that the coupling and
interaction of the {\it p}-mode waves with the magnetized
plasma can possibly cause the running waves.
More recently, it has been suggested that RPWs are slow low-$\beta$ waves propagating upwards along inclined magnetic field lines \citep{2006RSPTA.364..313B, 2007ApJ...671.1005B,2013ApJ...779..168J,2015AGUFMSH31B2414M} facilitating the propagation of non-thermal energy into the corona.
Some studies \citep{1992SoPh..138...93A,1998ASPC..155...49A,1996SoPh..167...79T, 2000A&A...355..375T,2003A&A...403..277R} 
show that they are waves originating from oscillating elements
inside the umbra. According to some studies, RPWs and umbral oscillatory events belong to the same traveling wave system and 
probably  the underlying  driving physical mechanisms are same. \citet{1998ApJ...497..464L} 
proposed that either the RPWs are driven by the umbral oscillatory events or they
share a common physical basis. The problem is that an
indisputable physical model  is not available which could attribute these two
phenomena to the same driving mechanism and also explain their
differences (periods of 3 and 5 min).

On the other hand, there  is  also evidence that umbral
oscillatory events of the chromosphere are not the source of RPWs
\citep{1972SoPh...27...71G,1975SoPh...41...81M,2000A&A...354..305C,2001A&A...375..617C,
2002A&A...381..279T,2004A&A...424..671K,2007ApJ...671.1005B}.
For example, \citet{2001A&A...375..617C} reported that RPWs are
more closely associated with photospheric umbral oscillations than
the chromospheric ones. \citet{2014ApJ...791...61F} observed RPWs within a solar pore and interpreted it 
as upwardly propagating  waves (UPWs). They found that the power enhanced at the boundary of the pore at about 3-5 min whereas in the chromosphere where the UPWs are observed, the power reduced.  Moreover, \citet{2004A&A...424..671K}
found that in most cases the running umbral waves terminate rather
abruptly at the umbral boundary and show no direct linkage with
RPWs. That means not all 3-min wave fronts can be traced out from
the umbra into penumbra. 
However, the question is where do the RPWs initially emerge in
chromosphere : is it from the inner umbra or umbral boundary of sunspots? 
Further, how are they linked with umbral oscillatory events - this remains an open question. Here, we present the first observations of RPWs linked to previously occuring RPWs and their further development into new RPWs.
We discuss, here, the study of the origin of emergence of RPWs from sunspots using the spectra of TiO, H$\alpha$, and 304 \AA{} for various atmospheric heights from the photosphere to lower corona. By employing time series analysis of imaging observations, we track umbral oscillatory events and their association with the preceding and following RPWs. The paper is organized as follows.
Section 2 details the observations and reduction of the data presented, Section 3 describes the analysis of the data and studies the umbral oscillatory events at different heights, and Section 4 summarizes and concludes.

\section{Observations and data processing} \label{sec:data}

High-resolution observations  were carried out of the leading sunspot of 
Active Region NOAA 12132, on August 5, 2014,  from New Solar Telescope 
\citep[NST,][]{2010AN....331..636C} operating at Big Bear Solar Observatory 
(BBSO).  We have employed the dataset already investigated by \citealt{2016ApJ...817..117S} to study the spiral structures of wavefronts in the sunspot. The sunspot of our observation is located at S09E08 as shown in Figure~\ref{fig:fig1}(a). 
Observations begun at 18:19 UT for a duration of 60 minutes. We used 
the broad-band filter imager of NST, with a field of view (FOV) of
70$\arcsec$ at 0.034$\arcsec$ pixel$^{-1}$ image scale to acquire continuum 
photospheric images every 15 s in TiO band (705.7 nm, 10 \AA{} bandpass).   
We  also employed  the Visible Imaging Spectrometer of NST that has a single Fabry-P\'{e}rot etalon to produce a narrow 0.07 \AA{} bandpass
over a 70$\arcsec$ circular FOV at 0.034$\arcsec$
pixel$^{-1}$ image scale. The chromospheric images were thus
acquired every 23 s by scanning  the H$\alpha$ spectral line from
its blue wing -1 \AA{} to red wing +1 \AA{} with  a step size of 0.2 \AA{}.
In addition, we also acquired the simultaneous space observations taken in 304
\AA{} line (formed in the transition and lower corona)  of
the Atmospheric Imaging Assembly on Solar Dynamics Observatory
\citep[SDO/AIA,][]{2012SoPh..275...17L}.
We chose the first H$\alpha-1.0$ \AA{} image as a
reference image to align all other images in this
passband. The relative shifts were recorded, and
used to register the images in the other passbands of
H$\alpha$.

Similarly, using the reference image, alignment was easily executed for
TiO images and one white-light image at 17:15 UT taken with
the Helioseismic and Magnetic Imager \citep[HMI,][]{2012SoPh..275..229S} 
on board SDO. The aligned white light image was then used to co-align the 
304 \AA{} images. Finally, Fast Fourier Transform (FFT) was applied to the
time-series images to generate the filtered component images, either 
in phase speeds of $v_{ph}>4$ km s$^{-1}$ (see \citealt{2016ApJ...817..117S}) 
or centering at certain frequency (e.g., 3.33 mHz, 5.55 mHz, etc).

\section{Analysis and discussion} \label{sec:analysis}
\subsection{ High and low frequency oscillations at different heights} \label{subsec:power}
\begin{figure}
\figurenum{1}
\epsscale{1.2}
\plotone{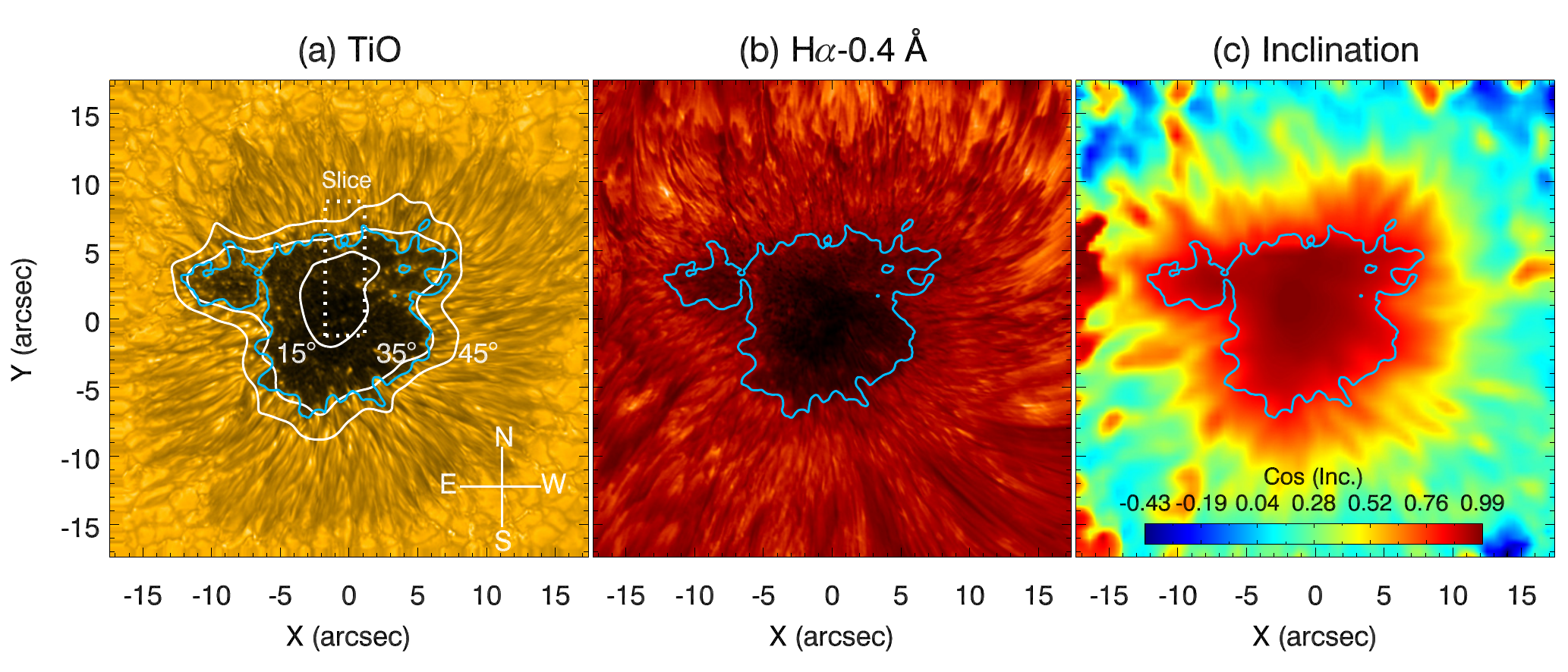}
\caption{Maps of the leading sunspot in active region NOAA 12132 on August 5, 2014. (a) is a TiO image, in which 
	a rectangle region marked by dotted lines is selected to be analyzed later. White contours 
	represent the inclinations of 15$\degr$, 35$\degr$ and 45$\degr$ and blue 
	contour in (a)$-$(c) denotes the umbral boundaries ($45\%$ of the maximum intensity). (b) is a H$\alpha-0.4$ \AA{} image. (c) 
	is a field inclination image. \label{fig:fig1}}
\end{figure}

\begin{figure}
    \epsscale{1.2}
    \figurenum{2}
    \plotone{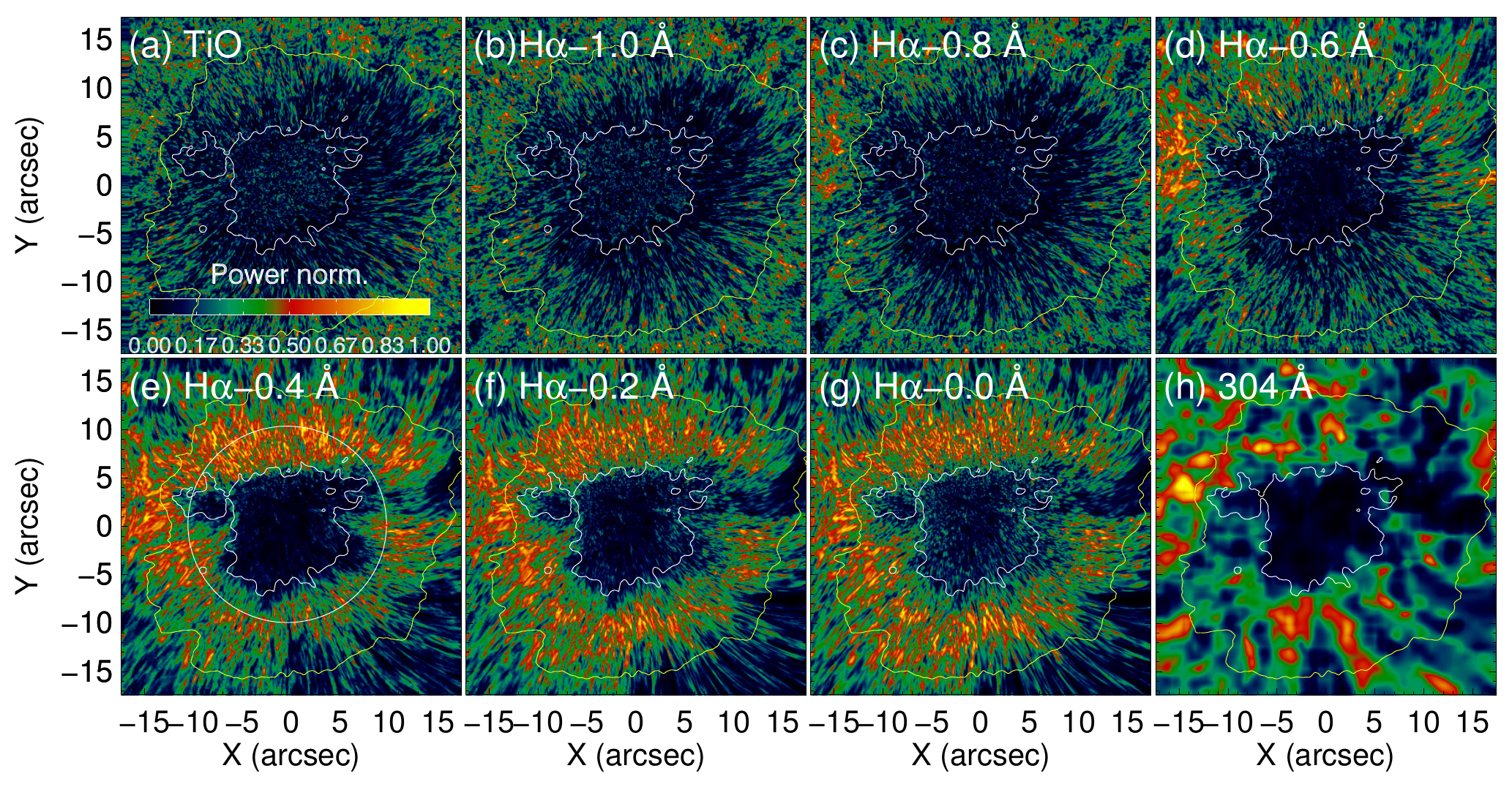}
    \caption {Power maps of the sunspot. (a)$-$(h) are spatial distribution of the normalized 
    	Fourier power for $3.8-8.0$ min oscillations taken in the indicated passband. Umbral 
    	regions of sunspots are shown in white contours and penumbral regions in yellow contours (see Figure~\ref{fig:fig1}). \label{fig:fig2}}
\end{figure}

\begin{figure}
	\epsscale{1.2}
	\figurenum{3}
	\plotone{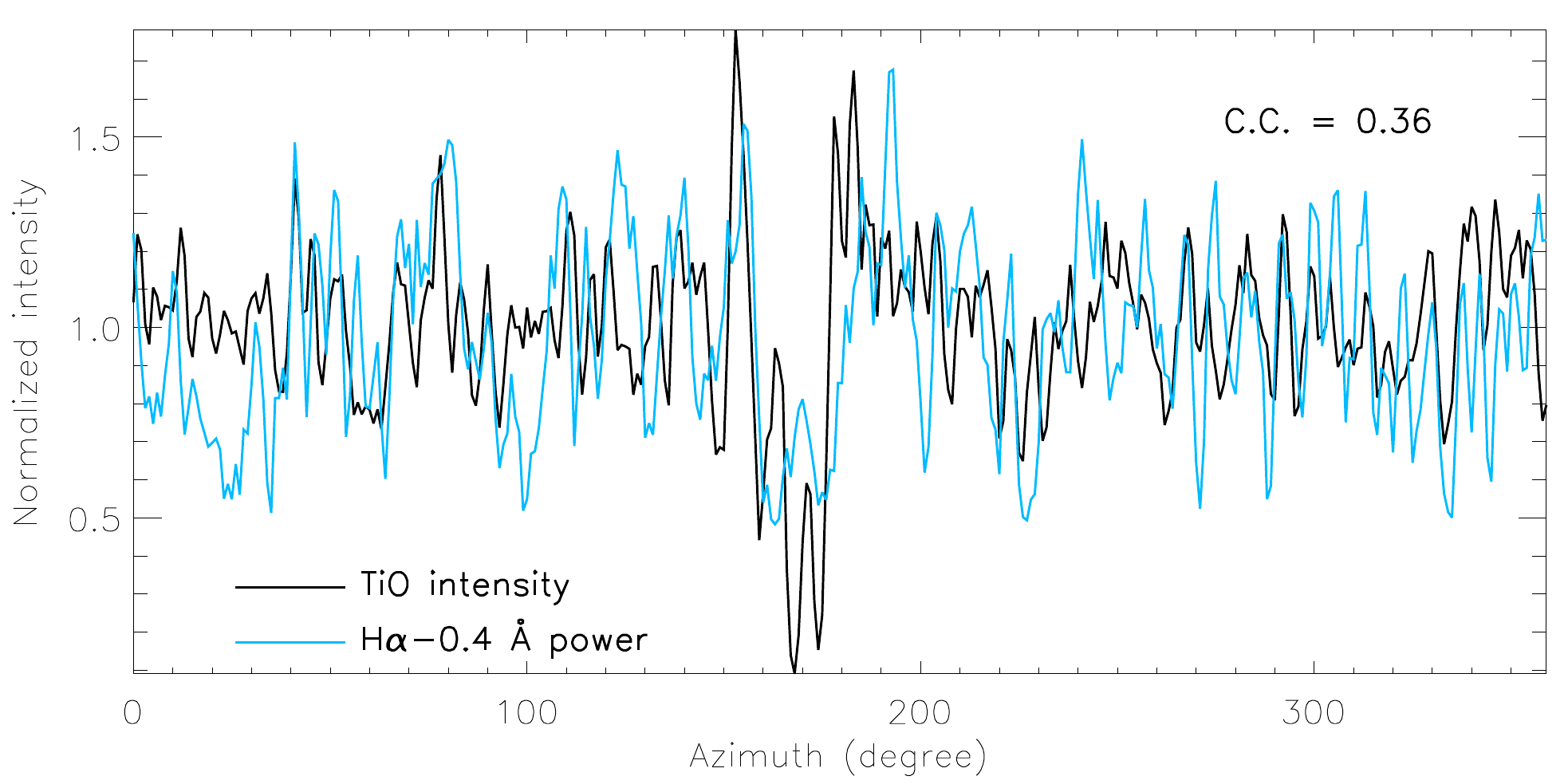}
	\caption {TiO intensity and  H$\alpha-0.4$ \AA{} oscillation power averaged over the circular slits shown in Figure~\ref{fig:fig2} .\label{fig:fig3}}
\end{figure}

 In Figure~\ref{fig:fig1}($a$), we show the map of sunspot in TiO image and its corresponding magnetic field inclination map  in Figure~\ref{fig:fig1}($b$) acquired on 2014-08-05. The magnetic field inclination of different range (15$\degr$, 35$\degr$ and 45$\degr$) is shown by different contours. We use these range of inclination to understand how the strength of oscillation varies from umbral center to umbral boundary.\\ Figure~\ref{fig:fig2} shows 
the power maps of $3.8-8.0$ min oscillations for the sunspot in the passbands of 
TiO, H$\alpha=1.0$, -0.8, -0.6, -0.4, -0.2, 00 \AA{} and 304 \AA{}, respectively. It is clear from Figures~\ref{fig:fig2} $(a)$ and $(b)$ that  the power is much weaker in the sunspot than in the quiet Sun  at  the photosphere.
However, the power in penumbrae 
becomes stronger with the increasing height (see Figures~\ref{fig:fig2} $(c)-(e)$). Evidently, even with
such high-resolution ($0.1\arcsec$) observations we still cannot confirm the 
existence of 5-min oscillations in the chromospheric umbrae (see Figure~\ref{fig:fig2} $(d)-(g)$). 
\citet{2013SoPh..287..149Z} suggested that resonant oscillations exist inside 
the sunspot, primarily inside the sunspot penumbra.  In adition to the overall power enhancement over the penumbra, there are tiny speckles of power scattered everywhere. These tiny speckles are associated much with the bright superpenumbral fibrils. We made a circular slit as shown in Figure~\ref{fig:fig2}($e$) and  plot the intensity curve for TiO and power of H$\alpha - 0.4$ \AA{} which is shown in Figure~\ref{fig:fig3}. It shows that there is probably, a slightly higher power concentration in the bright superpenumbral fibrils. We normalized the power spectra by adopting the method of division, wherein each point is divided with the value of maximum intensity inorder to compare the power spectra at different wavelengths and positions. The correlation coefficient for the two normalized intensity $\approx$ 0.36 showing a weak correlation between the intensity and power.

\begin{figure}
    \epsscale{1.2}
    \figurenum{4}
    \plotone{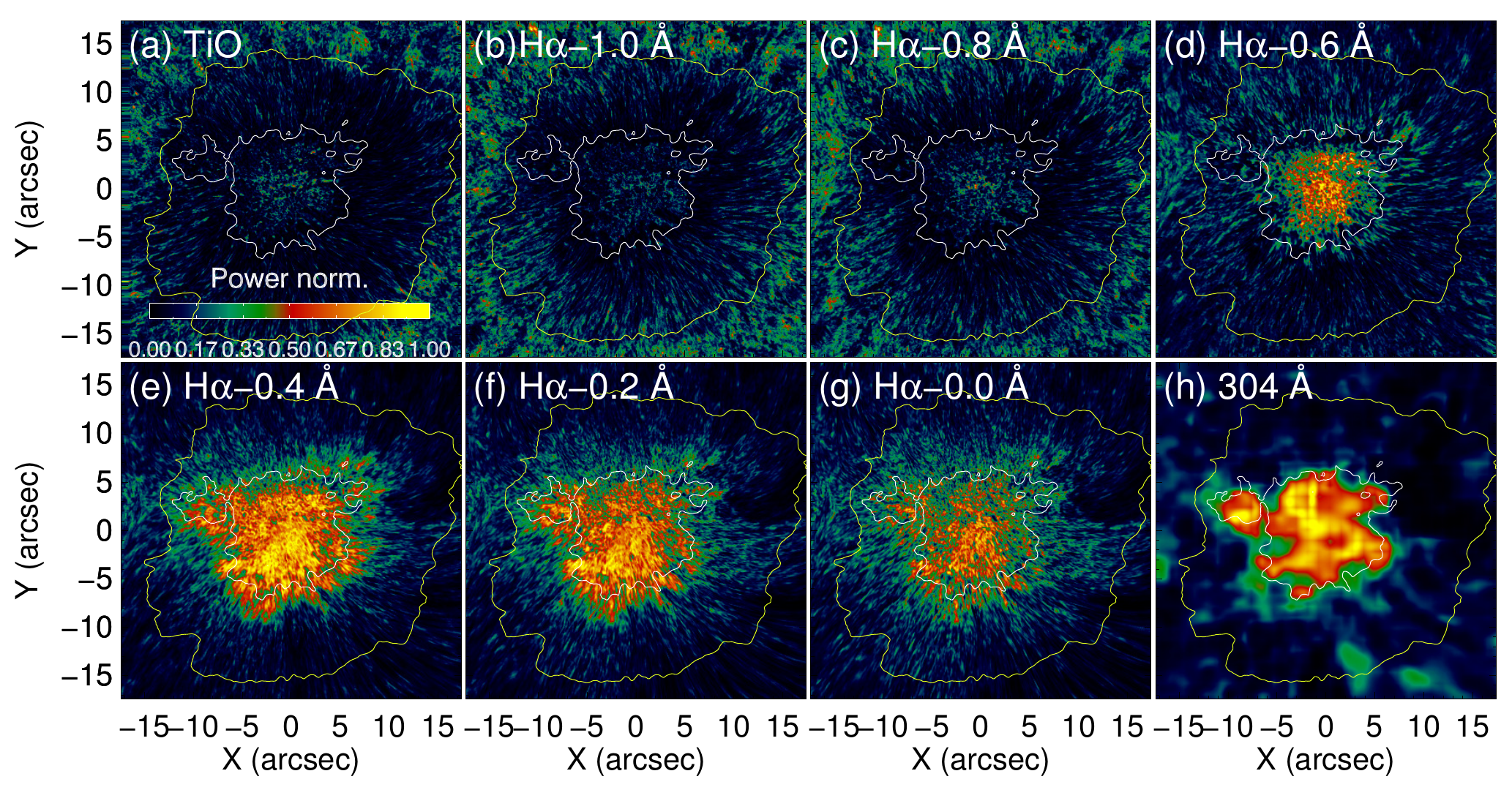}
    \caption {Similar to Figure 2 but for the higher frequency oscillations in a range of $1.7-3.8$ min.\label{fig:fig4}}
    
\end{figure}

\begin{figure}
	\epsscale{0.85}
    \figurenum{5}
    \plotone{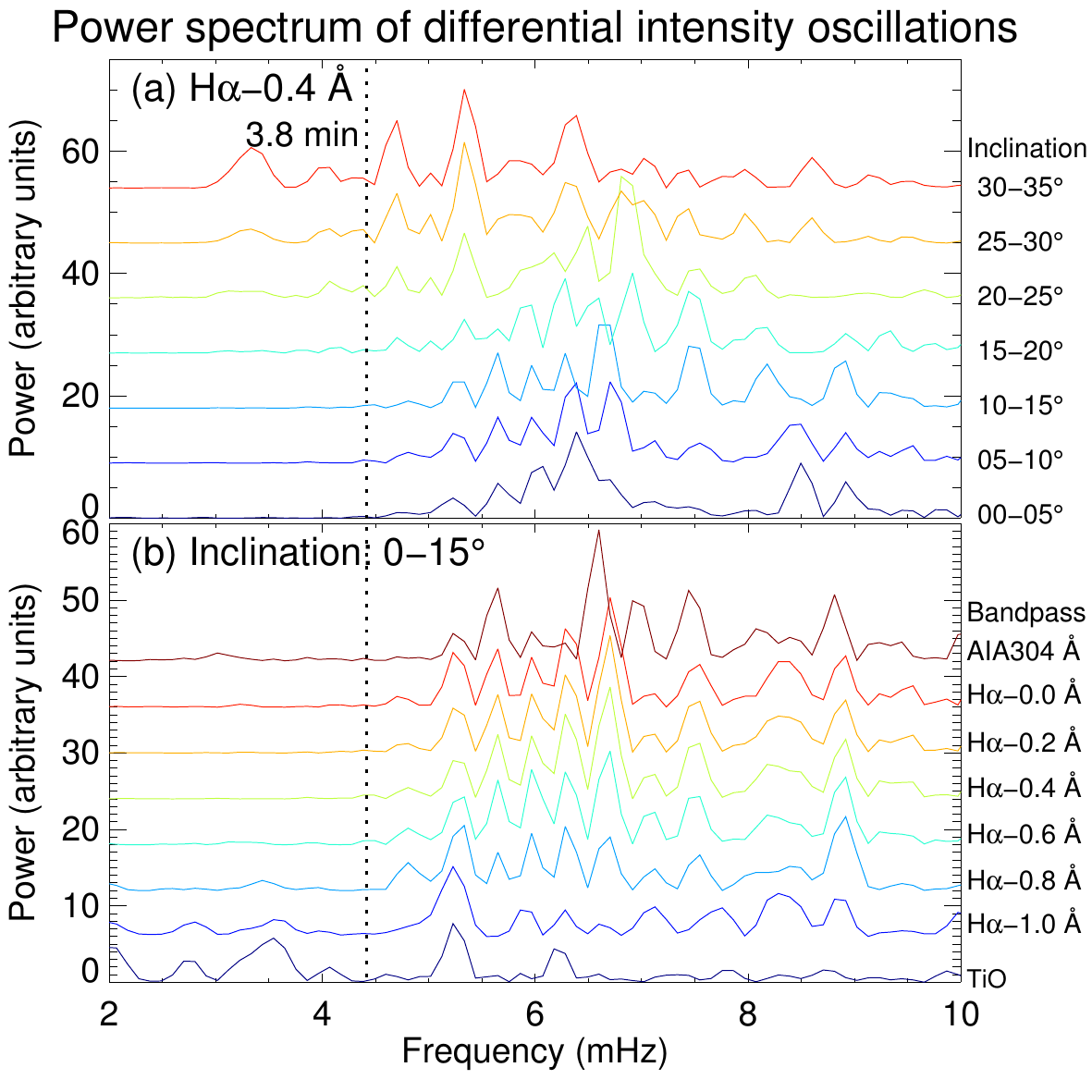}
    \caption{(a) Averaged power as a function of field inclination in the passband of 
        	H$\alpha-0.4$ \AA{}. (b) shows averaged power in a range of $0-15\degr$ as a 
        	function of passbands TiO, H$\alpha-1.0$ \AA{}, ... and 304 \AA{}. \label{fig:fig5}}
   
\end{figure}

Figure~\ref{fig:fig5} displays the power spectra of oscillations in 
the sunspot averaged over some  circular slits  with inclination in the range 
of 5 $\degr$ interval between 0$\degr$ and 35$\degr$ (e.g. $0-5\degr$, $5-10\degr$, upto $30-35\degr$) in the passband of 
H$\alpha - 0.4$ \AA{} (top panel), and averaged over a circular slit 
with $0-15\degr$ inclination in the passbands of TiO, H$\alpha - 1.0$, 
-0.8, ... and 304 \AA{} (bottom panel). The noise in the data is removed by performing numerical differentiation on all time series of intensity.  We concentrate more on low frequencies as low frequencies introduces less noise than the high frequencies. In H$\alpha$, beyond the photospheric penumbra, the high frequency power is almost undetectable while the low frequency power enhances. We find that most of the power in high-frequency oscillations (period = 1.7-3.8 min), is concentrated in the umbra (see Figure~\ref{fig:fig4}) whereas the power in low-frequency oscillations (period = 3.8-8 min) is concentrated in the penumbra.
We  arbitrarily take 3.8 min as the cut-off point of 3-min and 5-min band oscillations. 
In Figure~\ref{fig:fig5}$(a)$, 5-min oscillations show their 
strongest signal appearing in the inclination range of $30-35\degr$, which 
is close to umbral boundaries (see Figure 1). Subsequently, they decrease 
while approaching  the  umbral center and are  nearly undetectable in  the  range of $15-20\degr$. 
Figure~\ref{fig:fig5}$(b)$ show 5-min oscillations are stronger at the formation 
heights of TiO and H$\alpha-1.0$ \AA{}. However, they become nearly invisible
at the formation height of H$\alpha-0.8$ \AA{}. 

\begin{figure}
	\epsscale{1.0}
    \figurenum{6}
    \plotone{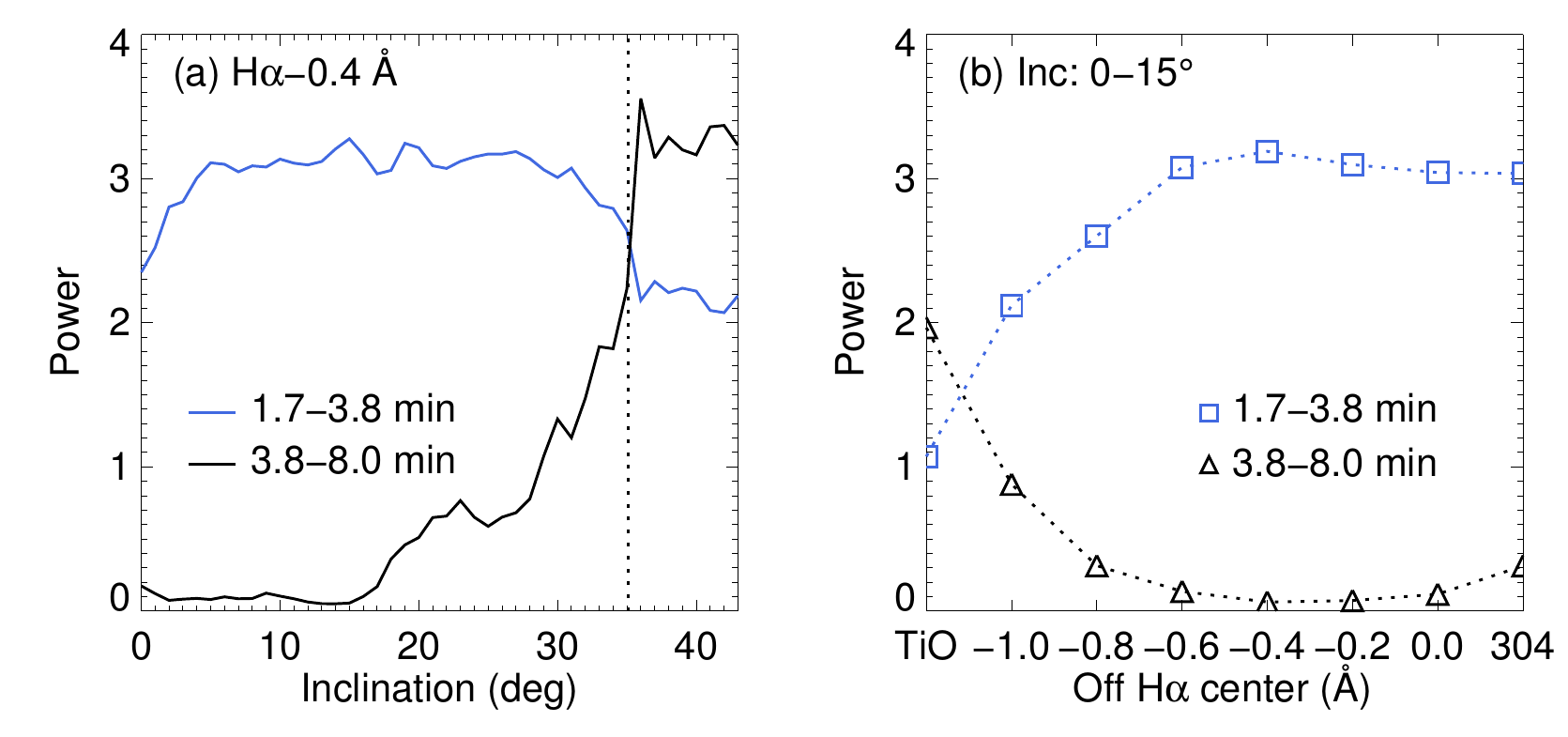}
     \caption{(a) Power spectra of H$\alpha-0.4$ \AA{} averaged over a circular slice with 
        	inclination range of $0-5\degr$, $5-10\degr$, ... and $30-35\degr$ in the sunspot, where the dotted-line gives the limits of the umbra and the penumbra. (b) shows spectra of 	TiO, H$\alpha-1.0$, ..., 0.0 \AA{} and 304 \AA{}, averaged over a circular slice with 	$0-15\degr$ inclination. For better visualization, each curve added with 9 starting from $5-10\degr$ line in (a) and 6 from H$\alpha-1.0$ \AA{} in (b). \label{fig:fig6}}
    
\end{figure}

We divided the power spectra in Figures \ref{fig:fig5}$(a)$ and \ref{fig:fig5} $(b)$ into 
two periodic ranges, 1.7 $< P <$ 3.8 min (short) and 3.8 $< P <$ 8 min (long), 
and the average power in the two ranges are shown in Figure \ref{fig:fig6}. 
The variation in power with inclination is  plotted in Figure~\ref{fig:fig6}$(a)$.
For the curve of $3.8-8$ min, the power increases exponentially with inclination 
in the range of $\sim15-36\degr$. Figure~\ref{fig:fig6}$(b)$  demonstrates the 
variation in power with height. For the $3.8-8$ min curve, it falls off 
exponentially with increase in height, and fades out while approaching the 
formation height of H$\alpha-0.4$ \AA{}. Generally, it is less than $5\%$ for the 
transmission rate of 5-min oscillation power from TiO to H$\alpha-0.4$, 
-0.2 and 0.0 \AA{}. Thus, it seems no 5-min {\it p }-mode waves can propagate 
vertically from photospheric to chromospheric umbra (e.g., within $15\degr$ 
inclination range). 
\subsection{Time- distance  diagram for umbral oscillatory events and RPWs} \label{subsec:diagram}
We constructed time-distance diagrams of the Doppler shift (difference of H$\alpha-0.4$ \AA{} 
and $+0.4$ \AA{}), H$\alpha-0.4$ \AA{} and AIA 304 \AA{} as seen in Figure~\ref{fig:fig7}, 
derived for the intensity averaged along a slit of $\sim3\arcsec$ width shown in Figure \ref{fig:fig1}. 
The connection between the events of umbral oscillations and RPWs are demonstrated in general. In terms of morphology, 
there is not much difference between panels (a) and (b) and they both show oscillations in 
umbral regions and wave propagations in penumbral regions. However, they are different from 
panel (c), that shows the oscillatory features in both umbra and penumbral regions. In the time-distance diagrams, we see a fork pattern forming around the umbral boundary in Figure~\ref{fig:fig7} $(a)$ and $(b)$. This pattern is similar to the fork pattern seen by \citet{2014ApJ...789..108C} but the explanation is quite different. We speculate that the formation of this fork pattern indicates that the umbral oscillatory event  emerges close to umbral boundary and propagates higher up and then a part of the wavefront segregates from the propagating wavefront as it reaches the umbral boundary and moves into the umbral center. This is explained in more detail in the following section.
To gather more information, we use a slit and produce two other diagrams in the two time 
intervals of 18:29:53$-$18:42:52 UT and 18:43:15$-$18:56:14 UT, 
respectively  as  shown  in Figure~\ref{fig:fig8}. In both the panels, there are 5 individual complete 
umbral oscillatory events, e.g., at 18:31:25 UT, 18:34:05 UT, etc (named after their  kick-off 
time), but corresponds to 4 RPWs in panel Figure~\ref{fig:fig8} $(a)$ and only 3 RPWs in panel 
Figure~\ref{fig:fig8} $(b)$, respectively, which might be due to the merging of some of the 
umbral oscillatory events together or may be it could not propogate at all. Another  puzzling feature 
is the association of some  events of umbral oscillations with their preceding RPWs. For example, events 
of 18:31:25 UT  and 18:34:05 UT were connected to their preceding RPW by some stripes, 
which indicates the inward propagation of  the associated  wavefront towards umbral center. 
In this paper, we report on 24 events of umbral oscillations in the time interval 18:19 to 
19:19 UT (as shown in Table 1). We then proceed to investigate in detail two of these events 
by employing time series of imaging observations.

\floattable
\begin{deluxetable}{ccCccccc}
	\tablecaption{ Umbral oscillatory events and RPWs in the period of 18:19$-$ 19:19 UT.}
	\tablecolumns{6}
	\tablenum{1}
	\tablewidth{0pt}
	\tablehead{
		\colhead{Umbral} &
		\colhead{Start time} &
		\colhead{$\theta$\tablenotemark{a}} & $\bar{\theta}$ &\colhead{Preceding\tablenotemark{b}} & \colhead{Following\tablenotemark{c}} & \colhead{$\bar{v}^{rpw}$}\\
		\colhead{oscillations} & \colhead{ (UT)} &
		\colhead{(degree)} & (degree)& \colhead{RPW} & \colhead{RPW}& \colhead{km s$^{-1}$	}
		}
	\startdata
	01 & 18:21:24 & $10-20\degr$ & $15\degr$&Yes & Yes & 9.8\\
	02 & 18:23:42 & $10-25\degr$ & $18\degr$&Yes & Yes$^\ast$& 9.8 \\
	03 & 18:25:40 & $10-30\degr$ & $20\degr$&Yes & Yes & 9.2\\
	04 & 18:27:58 & $15-25\degr$ & $20\degr$&Yes & Yes$^\ast$& 9.2\\
	05 & 18:31:25 & $40-30\degr$; $30-20\degr$; $15-25\degr$ &$35\degr$;25\degr$;20\degr$&Yes & Yes & 10.0\\
	06 & 18:34:05 & $15-30\degr$ & $23\degr$&Yes & Yes & 10.5\\
	07 & 18:36:23 & $10-25\degr$ & $18\degr$&Yes & Yes$^\ast$ & 10.5\\
	08 & 18:38:17 & $30-45\degr$; $10-30\degr$ & $38\degr$;$20\degr$&Yes & Yes & 13.8\\
	09 & 18:40:58 & $15-30\degr$; $30-45\degr$ & $26\degr$;$38\degr$&Yes & Yes & 12.8\\
	10 & 18:44:01 & $25-40\degr$; $30-45\degr$ & $26\degr$;$38\degr$&Yes & Yes & 9.7\\
	11 & 18:46:18 & $10-20\degr$ & $15\degr$&Yes & Yes & 8.0\\
	12 & 18:49:22 & $15-30\degr$ & $23\degr$&Yes & Yes & 7.5\\
	13 & 18:51:39 & $10-20\degr$ & $15\degr$&Yes & Yes$^\ast$ & 7.5\\
	14 & 18:54:19 & $0-15\degr$ & $8\degr$&? & Yes & 10.0\\
	15 & 18:57:22 & $25-40\degr$ & $33\degr$&Yes & Yes & 8.5\\
	16 & 19:00:26 & $25-40\degr$;$10-25\degr$ &$33\degr$;18$\degr$& Yes & Yes$^\ast$ & 8.5\\
	17 & 19:03:09 & ? & ? &? & Yes & 6.0 \\
	18 & 19:05:50 & $25-40\degr$ & $33\degr$&Yes & Yes & 14\\
	19 & 19:08:08 & $25-35\degr$ & $30\degr$&Yes & Yes & 10\\
	20 & 19:10:03 & $10-25\degr$ & $18\degr$&Yes  & Yes$^\ast$ & 10 \\
	21 & 19:11:57 & $10-30\degr$ & $20\degr$&Yes & Yes  & 12\\
	22 & 19:13:52 & $0-35\degr$ & $18\degr$&Yes & Yes$^\ast$ & 12\\
	23 & 19:16:09 & $0-10\degr$ & $5\degr$&No  & Yes   & 9.0\\
	24 & 19:19:13 & $25-35\degr$ & $30\degr$&Yes & ? & ?\\
	\enddata
	\tablenotetext{a}{Range of magnetic field inclination at the initial emergence of the umbral oscillatory event.}
	\tablenotetext{b}{Whether was the event related to its preceding RPW?}	
	\tablenotetext{c}{Whether did the event develop into the following RPW?}	
	\tablecomments{Symbol $\ast$ denotes the following wavefront catching up and merging with its preceding one.}
\end{deluxetable}

\subsection{Imaging observations of umbral oscillatory events and RPWs} \label{subsec:ima}

We investigate upon two typical umbral oscillatory events starting at 18:31:25 UT in panel $(e)$ of Figure~\ref{fig:fig9} and at 18:51:39 UT in panel $(d)$ of Figure~\ref{fig:fig10}, respectively. For the first event, the three dark patches, A, B and C  highlighted by red squares appeared initially close to/on umbral boundaries. It has central field inclinations  of $\sim35\degr$ (A), $25\degr$ (B) and $20\degr$ (C)  respectively (see Figures~\ref{fig:fig9}$(b)$ and $(c)$). A and B patches separated  off from the umbral boundaries (see white arrows in panel Figure~\ref{fig:fig9} $(c)$) and patch C stayed where it was. The three patches become enhanced at 18:31:02 UT and then shrink towards the umbral center (see the movie). At 18:31:25 UT the first event of umbral oscillation began as shown in Figure~\ref{fig:fig8} $(a)$.

Later, a dark circular patch formed at 18:31:48 UT in umbral center and the wavefront began 
to expand at 18:32:10 UT. In the meantime, the propagation in clockwise direction at the top takes a spiral form along the trajectory of wavefront (See Figure~\ref{fig:fig9}$(g)-(i)$).  In Figure~\ref{fig:fig9} $(i)$,  the spiral's top end marked in a red square (denoted as A) separated from its main part and pushed itself towards the umbral center (see red circles in Figure~\ref{fig:fig9} $(j)$) and a new oscillatory event began again (see Figure~\ref{fig:fig9}  and ~\ref{fig:fig8}). We also notice that the spiral's tail which has moved to umbral boudaries made unticlockwise motions along the boundaries as shown by arrows in Figure~\ref{fig:fig9} $(h)$ and $(i)$. At 18:34:05 UT we mark it by a red square (denoted as B) in Figure~\ref{fig:fig9}$(j)$ and at next time, it suddenly dived into umbral center to merge with patch A. From then on, the merged patch expanded in radial direction (see Figure~\ref{fig:fig9}$(j)-(l)$)). Also, it is immediately noticeable that the main part of the preceding wavefront crossed the umbral boundaries and became a circular trajectory of RPW.  

It is seen in Figure~\ref{fig:fig8} that the stripes are stacked reversely, slightly slanted from 18:33:42 UT to 18:35:14 UT suggesting an inward propagation of wavefronts. 
Figure~\ref{fig:fig9}$(i)$ and $(j)$ clearly show the spiral's top end propagated downward which provided evident proof for this explanation. It is interesting that the stripes preceding and following the above ones also show this slanted feature.   

 Similarly, the second event occurring at 18:51:39 UT and in the period of 18:43:15-18:56:14 UT visualized in Figure~\ref{fig:fig8} and Figure~\ref{fig:fig10} showed similar behavior wherein the wavefront propagate towards the umbral center and merges with the preceding wavefront. A part of the merged wavefront jumps into the umbral center forming a new wavefront which ultimately results into a new umbral oscillation while the remaining wavefront crosses umbral boundaries to form RPW.  

In summary, we have interpreted the developments of features of dark patches A or B in the above 5 events with the wavefront initially reaching umbral boundaries without showing any further propagation along radial and azimuthal directions. Then, it begins to propagate towards the umbral center the next moment.  Hence, 
it appears to build a connection between the preceding and the following running waves. Some of the  major features of umbral oscillatory events and RPWs are summarized in Table 1. It is important to emphasize that the wavefronts in some events were having muti-spiral structures  and are complicated to determine their initial emergence, whether related to the preceding RPW or not, e.g., event 17 in the table . 

\subsection{Dominant oscillatory frequency in the sunspot umbra} \label{subsec:dom}
The distribution of dominant oscillating 
frequency in the sunspot umbra for event 18:51:39 UT is shown in three phases (see Figure~\ref{fig:fig11}$(a)-(c)$): The initial emergence followed by its propagation towards umbral center and finally its development into RPW.  The power of short period oscillation was weak at umbral center, before and after the wavefront propogation (see Figure~\ref{fig:fig11} $(a)-(b)$).  Also,  one can see that the dominant periods are very much different along the wavefront direction, and the averaged periods over the wavefront edges in 
the panels  are $2.7\pm5.3$ min, $2.8\pm5.9$ min and $4.0\pm7.0$ min, respectively. This may indicate that the running waves in the sunspot are broadband waves. 

\subsection{Active Region NOAA 12127} \label{sec:one}

We find similar phenomena occurring in a sunspot of AR NOAA AR 12127. This AR has a  complicated morphology due to its light bridges. Figure~\ref{fig:fig12} $(b)-(d)$ show power maps of $3.8-8.0$ min oscillations of the sunspots in the passbands of TiO, H$\alpha-0.4$ \AA{} and H$\alpha$-line center.  The complicated morphology of the object may impede the revealing of the general regularities and patterns in the observed phenomenon. Inspite of this, it is clearly seen in Figure~\ref{fig:fig13} that the wavefront emerged at the center of the umbra, then expanded and rotated clockwise. The wavefront propagates in both azimuthal and radial directions. It takes a spiral form and moves towards the umbral center and a new oscillatory event begins similar to the evnts in AR NOAA 12132.

\section{Conclusions} \label{sec:cons}
\begin{figure}
    \figurenum{7}
    \epsscale{0.85}
    \plotone{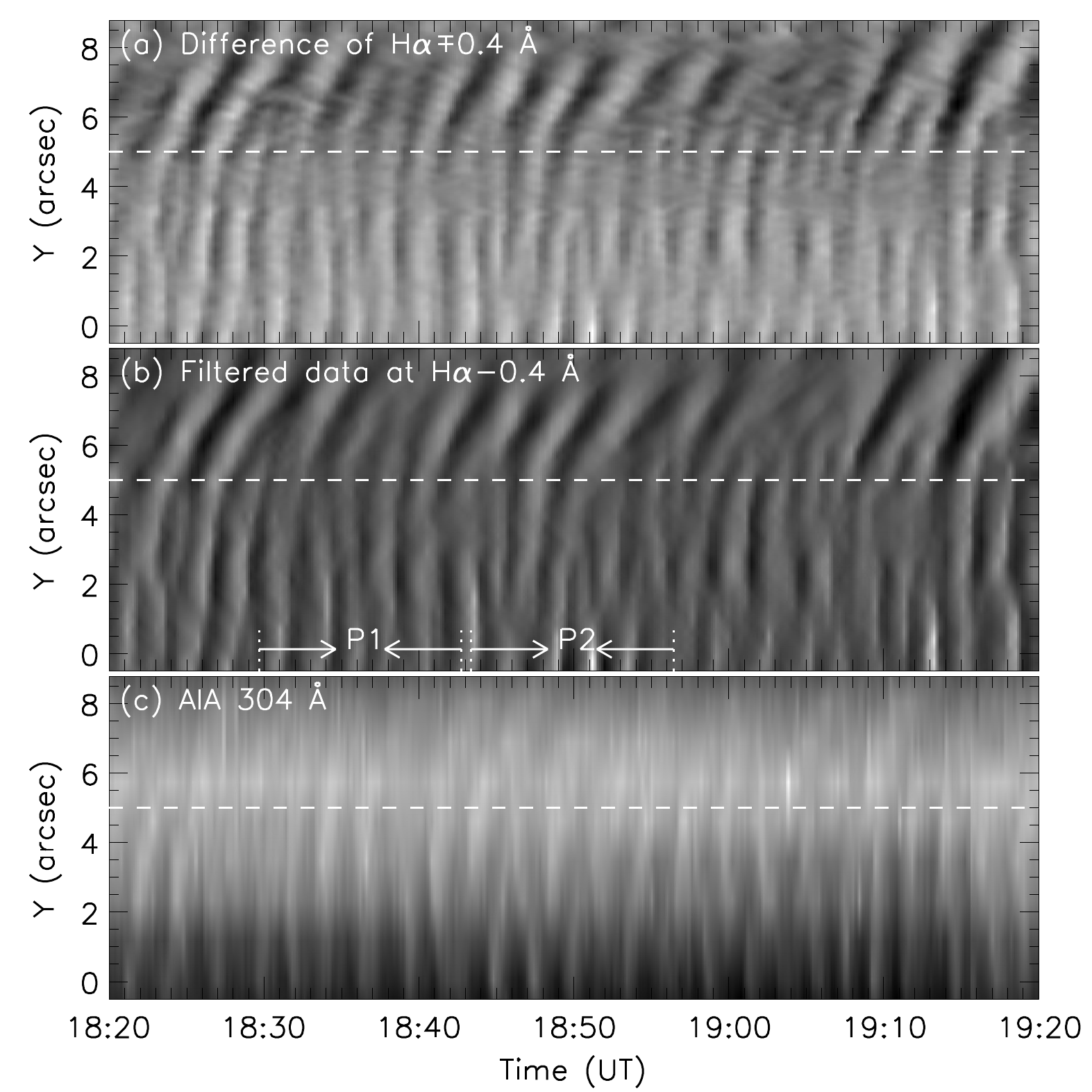}
    \caption{Time-distance diagrams of the Doppler shift (difference of H$\alpha\mp0.4$ \AA{}), 
        		H$\alpha-0.4$ \AA{} and AIA 304 \AA{}, derived for the intensity averaged along the width 
        		of the slice shown in Figure~\ref{fig:fig1} for the entire time sequence. White dashed line 
        		marks umbral boundary. The data in (b) marked by the arrows are to be analyzed in the 
        		following figure. } \label{fig:fig7}
    
\end{figure}

\begin{figure}
    \epsscale{0.9}
    \figurenum{8}
    \plotone{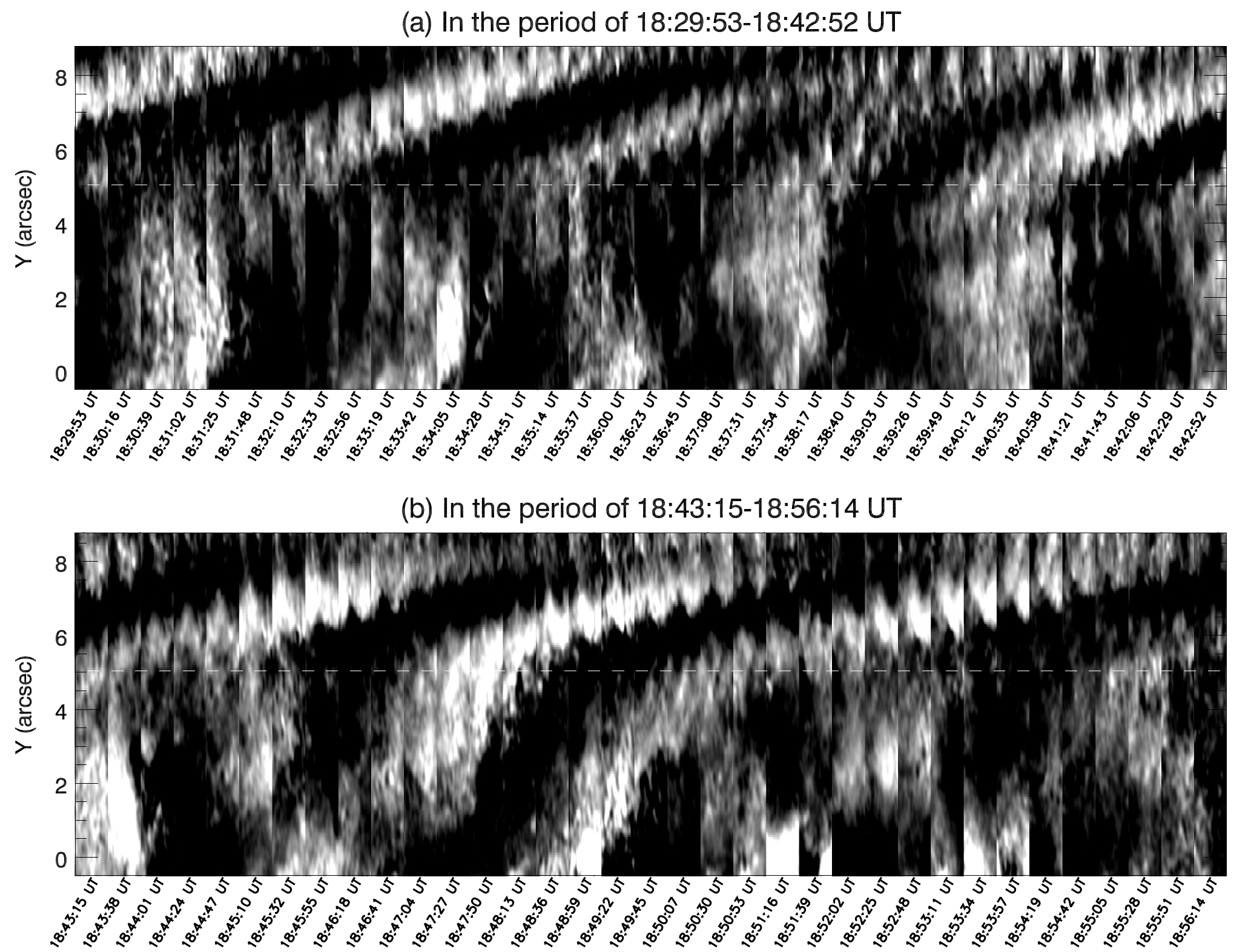}
    \caption{Time-distance diagrams for the slice shown in Figure~\ref{fig:fig1} within the periods of 
        	18:29:53$-$18:42:52 UT (a) and 18:43:15$-$18:56:14 UT (b). White dashed line marks umbral 
        	boundary.\label{fig:fig8}}
   
\end{figure}

\begin{figure}
   \epsscale{1.2}
	\figurenum{9}
	\plotone{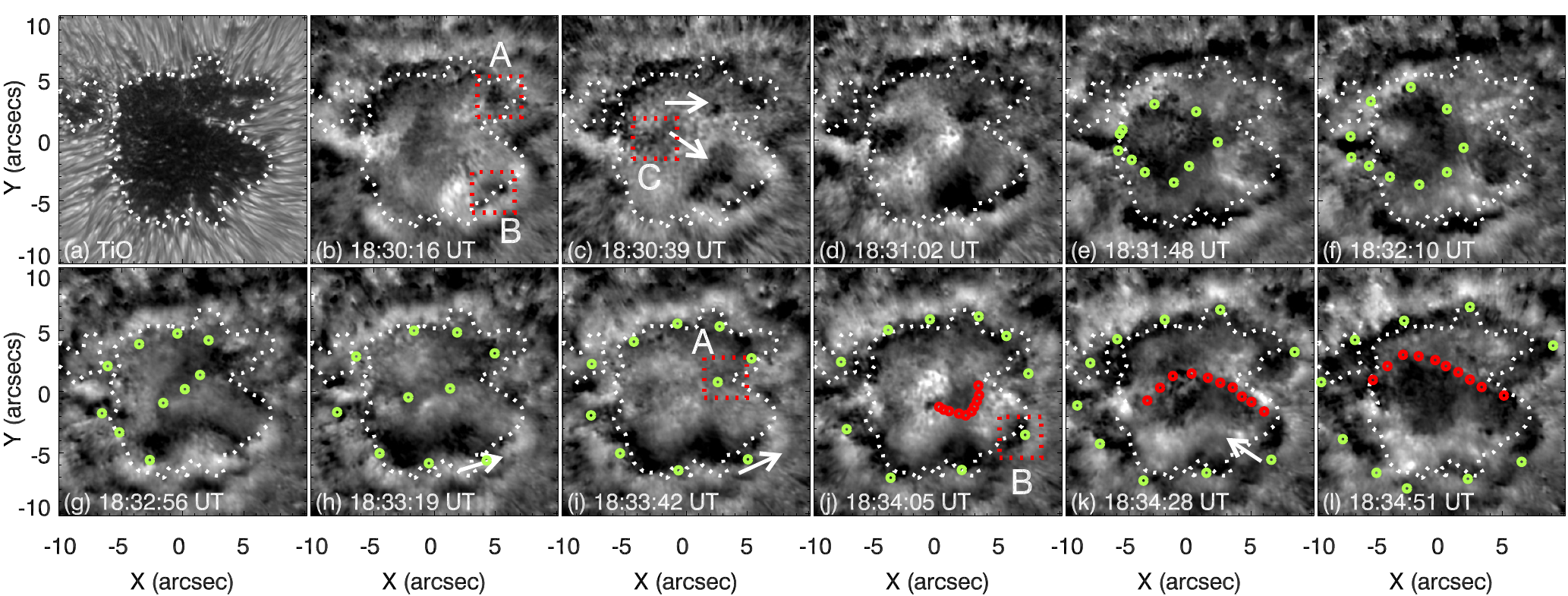}
	 \caption{$(a)$ is a TiO map for reference and white dotted contours in it and the other panels mark umbral boundaries. $(b)-(l)$ are the time series of filtered H$\alpha-0.4$ \AA{} images with phase speeds $>4$ km s$^{-1}$, on which circles are superposed to highlight the trajectories of running wavefronts.  Red squares and capital letters A,B and C mark initial emerging locations of the next  umbral oscillatory event. An
	 animated movie is available online \label{fig:fig9}}
	
\end{figure}

\begin{figure}
\epsscale{1.2}
	\figurenum{10}
	\plotone{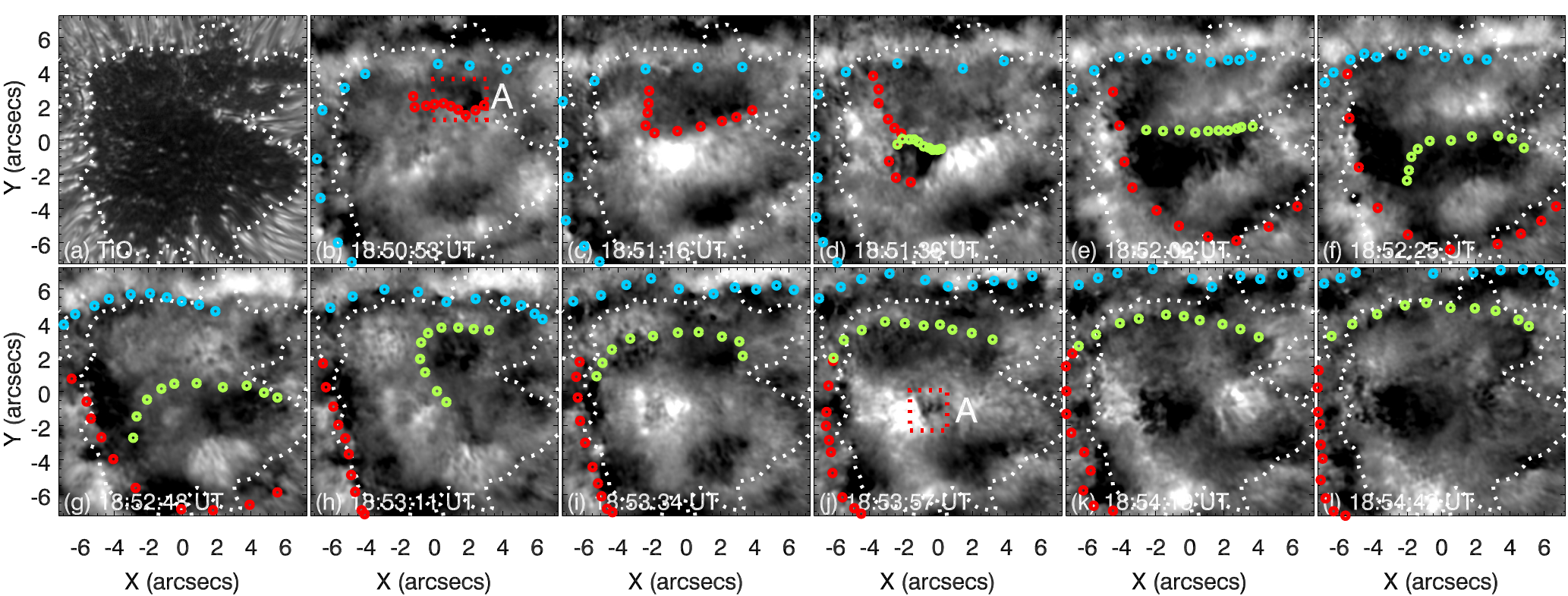}
	\caption{Similar to Figure 9, but for umbral oscillatory events 
		of 18:51:39 UT and 18:53:57 UT. \label{fig:fig10}}

\end{figure}

\begin{figure}
	 \epsscale{1.2}
    \figurenum{11}
    \plotone{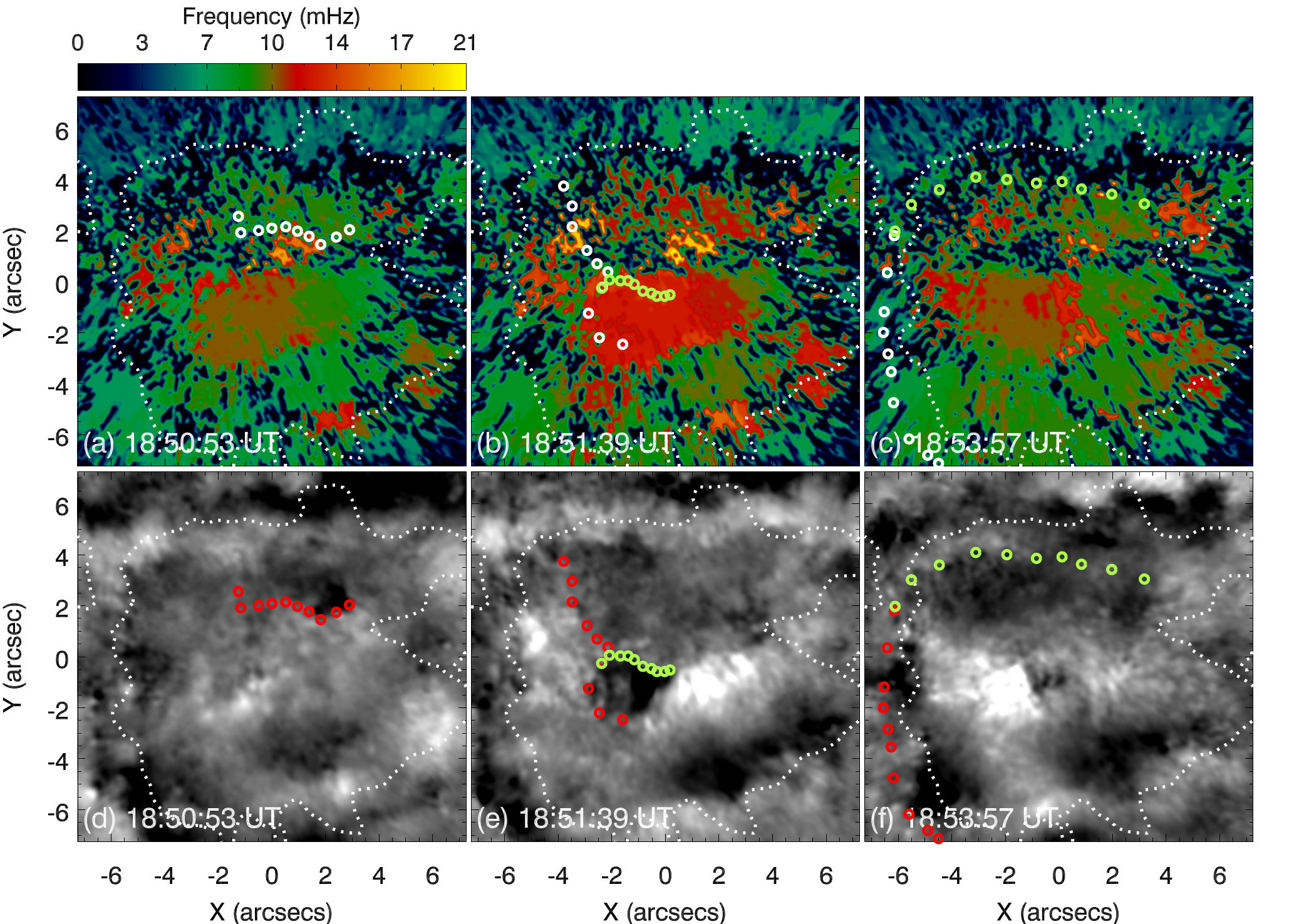}
    	\caption{$(a)-(c)$ are distributions of dominant oscillatory frequency in the sunspot umbra, 
    	 obtained with wavelet analysis for the time series of H$\alpha-0.4$ \AA{}
             filtered images.  $(d)-(e)$ are the corresponding H$\alpha-0.4$ \AA{} filtered images for event 
             of 18:51:39 UT in the three phases, first emerging, propagating to umbral center and developing 
             into RPW highlighted with circles. White contours in all panels denote umbral boundaries. \label{fig:fig11}}
    
\end{figure}

\begin{figure}
	\epsscale{1.2}
    \figurenum{12}
    \plotone{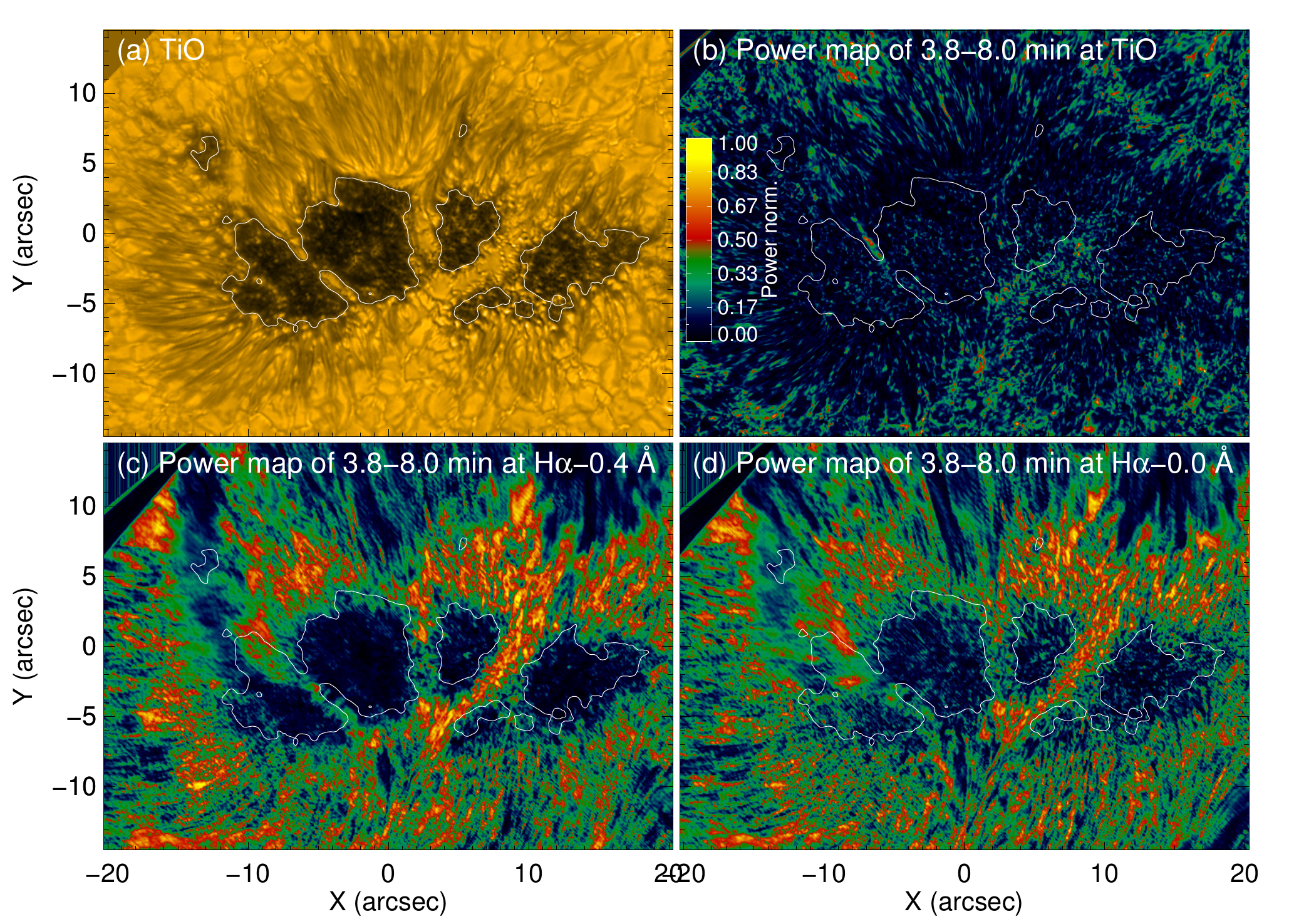}
    \caption{ Maps of main sunspots of AR 12127. (a) show the location of sunspots in TiO image. $(b)-(d)$ are corresponding spatial distribution of normalized Fourier power for $3.8-8.0$ min oscillations taken in the indicated passband. Note that umbral regions of sunspots are shown in white contours.  \label{fig:fig12}}
   
\end{figure}

\begin{figure}
	\epsscale{1.2}
    \figurenum{13}
    \plotone{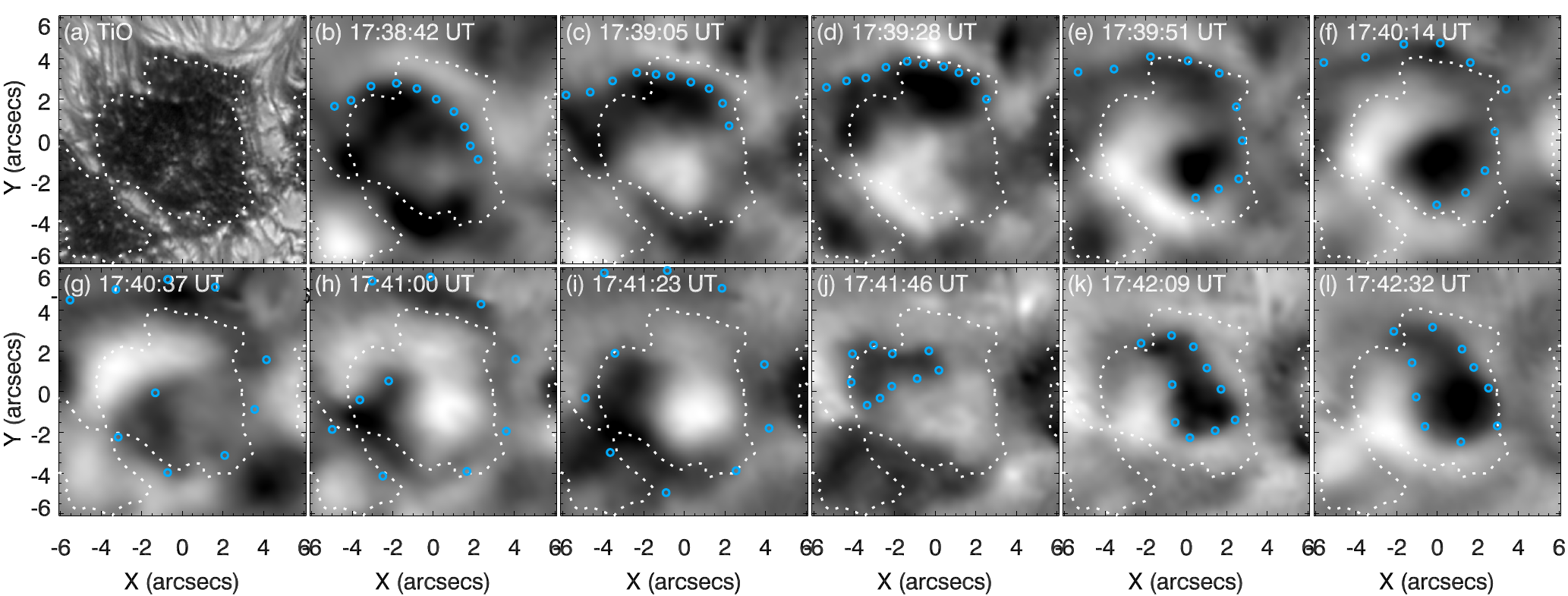}
 \caption{$(a)$ is a TiO map for reference and the white dotted contours marks the umbral boundaries. This shows the temporal evolution of the filtered H$\alpha-0.4$ \AA{} images with phase speeds $v_{ph}> 4$ km s$^{-1}$. Blue dots show the directions of wave propagation.  \label{fig:fig13}}
 \end{figure}

We have performed high-resolution imaging observations of active region NOAA AR 12132 to investigate umbral oscillatory events and RPWs. The main results are  the following: The long period (e.g., $3.8-8.0$ min) oscillations are hardly detectable within 
$15\degr$ field inclination range in  the  chromospheric umbra  which is consistent with the theory of MHD waves propagating in a simple, vertically gravitationally stratified atmosphere. The power observed in passbands of H$\alpha-0.4$, H$\alpha-0.2$ and H$\alpha$  line center  is  only  $5\%$ of that observed in TiO, suggesting that there is only $5\%$  
transmission rate for 5-min oscillation power from TiO to H$\alpha-0.4$, -0.2 and 
line center. Moreover, with imaging filtered observations we find that most of the umbral 
oscillations initially emerges  either at or close to umbral boundaries and also are a part of the preceding RPWs. This new result is also consistent with MHD wave propagation in inclined magnetic field embedded in gravitationally stratified plasma,  These fractional wavefronts are found to be separated from the preceding 
wavefronts and moves into  the  umbral center, where they expand radially and begins as a new 
oscillation  (An animation showing this phenomena is available online). A closer look at Figures~\ref{fig:fig7}$(f)$ reveals that the new wavefront emerges at the umbral center. The  panels in Figures~\ref{fig:fig7}$(a)-(e)$ shows that the umbral oscillatory events are related to the preceding RPWs whereas the panels in Figures~\ref{fig:fig7}$(g)-(l)$ shows that the umbral oscillatory events are related to the following RPWs.  In this way, nearly all umbral oscillatory events connect to the earlier occuring RPWs and also develop into new RPWs . This kind of connection between the umbral oscillatory events and RPWs is reported for the first time through this work. Also, along the wavefront edges, the period of running waves show a large spread. We finally remark that our new results contribute to an observational evidence about how an umbral oscillation is associated with a preceding and following RPWs.

Figures~\ref{fig:fig7} and ~\ref{fig:fig9} clearly demonstrates that the wavefronts form a   spiral structure suggesting the waveguides to be twisted  \citep[see, e.g.][]{2010ApJ...722L.194}. This observational fact might be the reason why we first see wavefront at high latitudes and then at lower latitudes at umbral center. However, 
the problem is still not resolved: RPWs being closely associated with  the events of  umbral oscillation, 
but outside the umbral boundaries 5-min signals go undetected. It has been interpreted earlier that on highly inclined magnetic flux tubes, with the cutoff period increase 
generates magnetoacoustic portals for the propagation of long$-$period magnetoacoustic waves in the chromosphere
which is also referred to as the leakage of {\it p}-modes or photospheric oscillations  \citep[see, e.g.][]{2004Natur.430..536D} into the chromosphere. It is worth mentioning that through MHD simulations \citet{2008ApJ...676L..85K} demonstrates that 5 minute oscillations can leak into the chromosphere through small-scale vertical magnetic flux tubes due to the efficiency of energy 
exchange by radiation in the solar photosphere that can lead to a significant reduction 
of the cutoff frequency and may allow the propagation of the 5 minute waves 
vertically into the chromosphere. We interpret that,  to a certain extent,  this mechanism strongly supports our observation.

\section{Acknowledgments}

This work is supported by National Basic Research Program of China under grant 11427901, 11773038, 11373040, 11303052 and 11303048. BBSO operation is supported by NJIT, US NSF AGS-1250818 and NASA NNX13AG14G, and NST operation is partly supported by the Korea Astronomy and Space Science Institute and Seoul National University, and by strategic priority research program of CAS. T.G.P would like to acknowledge the financial support from CAS-TWAS Presidents PhD fellowship - 2014. R.E is grateful to the Science and Technology Facilities Council (STFC) UK and The Royal Society, UK. R.E also acknowledges the support received by the CAS Presidents International Fellowship Initiative, Grant No. 2016VMA045, and the warm hospitality received at NAOC of CAS , Beijing, where part of his contribution was made. We thank referee for the review and for useful comments and suggestions.

\bibliographystyle{apj}
\bibliography{ms}

\listofchanges

\end{document}